\documentclass[twocolumn, showpacs,preprintnumbers,amsmath,amssymb]{revtex4}
\usepackage{amssymb}
\usepackage{xcolor}
\usepackage{graphicx}
\usepackage{dcolumn}
\usepackage{bm}
\usepackage{multirow}
\usepackage{booktabs}
\usepackage{natbib}
\usepackage{soul}
\usepackage{amssymb}
\usepackage{amsmath}	
\usepackage{color}
\usepackage{cases}

\begin{document}
\preprint{APS/123-QED}

\title{Reactive collisions between electrons and BeH$^+$ above dissociation threshold}

\author{E. Djuissi$^{1}$}
\author{J. Boffelli$^{1}$}
\author{R. Hassaine$^{1}$}
\author{N. Pop$^{2}$}\email[]{nicolina.pop@upt.ro}
\author{V. Laporta$^{3}$}
\author{K. Chakrabarti$^{4}$}
\author{M. A. Ayouz$^{5}$}
\author{J. Zs Mezei$^{1,6}$}\email[]{mezei.zsolt@atomki.hu}
\author{I. F. Schneider$^{1,7}$}\email[]{ioan.schneider@univ-lehavre.fr}
\affiliation{$^{1}$LOMC-UMR6294, CNRS, Universit\'e Le Havre Normandie, 76600 Le Havre, France}%
\affiliation{$^{2}$Department of Physical Foundation of Engineering, University Politechnica of Timisoara, 300223, Timisoara, Romania}%
\affiliation{$^{3}$Istituto per la Scienza e Tecnologia dei Plasmi, CNR, 70126 Bari, Italy}%
\affiliation{$^{4}$Dept. of Mathematics, Scottish Church College, 700006 Kolkata, India}%
\affiliation{$^{5}$LGPM, CentraleSup\'elec, Universit\'e Paris-Saclay, 91190 Gif-sur-Yvette, France}%
\affiliation{$^{6}$Institute for Nuclear Research (ATOMKI), H-4001 Debrecen, Hungary}%
\affiliation{$^{7}$LAC-UMR9188, CNRS Universit\'e Paris-Saclay, F-91405 Orsay, France}%
\date{\today}

\begin{abstract}
Our previous studies of dissociative recombination, and vibrational excitation/de-excitation of the BeH$^+$ ion, based on the multichannel quantum defect theory, are extended to collision energies above the dissociation threshold, taking into account the vibrational continua  of the BeH$^{+}$ ion and,  consequently, its dissociative excitation.  We have also significantly increased the number of dissociative states  of $^{2}\Pi$, $^{2}\Sigma^{+}$ and  $^{2}\Delta$ symmetry included in our cross section calculations,  generating the most excited-ones by using appropriate scaling laws. Our results  are suitable for modeling the kinetics of BeH$^{+}$ in edge fusion plasmas for collision energies up to 12 eV.
\end{abstract}


\maketitle

\section{Introduction}

Beryllium is considered as plasma-facing material for the International Thermonuclear Experimental Reactor (ITER)~\cite{ITER}. Therefore, understanding its release and its transport is a crucial aspect of the project. 
The beryllium wall will be exposed to plasma heat radiation and particle bombardment at the edge of the reactor inducing chemical erosion~\cite{Reiter2012, Phil:06}. Consequently, the beryllium atoms will enter the plasma and can form, by reaction with the fuel atoms (H, D, T), molecular species like BeH, BeD, and BeT, and their ions in fusion device~\cite{Reiter2012}. Other molecular species will also be formed as the beryllium interacts with impurities. The interaction of the edge and divertor hydrogen plasma with the  beryllium walls results in the formation and release of  beryllium hydrides. The lower temperatures in the edge and divertor plasma ($\sim$0.5–100 eV)  facilitate a large spectrum of collision processes and reactions between these species and the main plasma constituents of these regions. 

While the collision processes at the incident electron energy below the ionic dissociation threshold - i.e. 2.7 eV - have been the subject of several previous studies~\cite{Niyonzima2013,Niyonzima, Laporta2017, bed, Pop2021}, the extension of the collisional processes to higher energies - above the ionic dissociation threshold - is the main motivations of the present study. This is essential for the kinetic modeling of beryllium transport in the plasma and the determination of influxes of beryllium hydrides into the plasma~\cite{Kalyan2012}.  
 
Moreover, BeH and its cation have been identified in stars including the sun, in sunspots~\cite{Gaur1973}, and in several comets~\cite{Sauval1984,shabara2006}. Therefore, the accurate quantification of the relevant elementary processes involving these molecular species is of great importance for determining with the highest precision the physical and chemical conditions of these astrophysical media.

One of the most important reactive processes that can significantly change the character of the plasma is dissociative recombination (DR):
\begin{equation}\label{eq:DR}
 \mathrm{BeH}^{+} +\mathrm{e^{-}} \longrightarrow  \mathrm{Be + H}
\end{equation}
assisted by vibrational transitions (VT), vibrational excitation (VE) and de-excitation (VdE):
\begin{equation}\label{eq:VT}
 \mathrm{BeH}^{+}(v_{i}^{+}) +\mathrm{e^{-}} \longrightarrow  \mathrm{BeH}^{+}(v_{f}^{+})+\mathrm{e}^{-}
\end{equation}
and, at higher collision energies, by dissociative excitation (DE) :
\begin{equation}\label{eq:DE}
 \mathrm{BeH}^{+}+\mathrm{e}^{-}\longrightarrow  \mathrm{Be}^{+}+\mathrm{H}+\mathrm{e}^{-}.
\end{equation}
where $v_{i}^{+}$ and $v_{f}^{+}$ are the initial and final vibrational quantum numbers of the ion.

 \begin{figure*}[t]
      \centering 
      \includegraphics[width=0.75\linewidth]{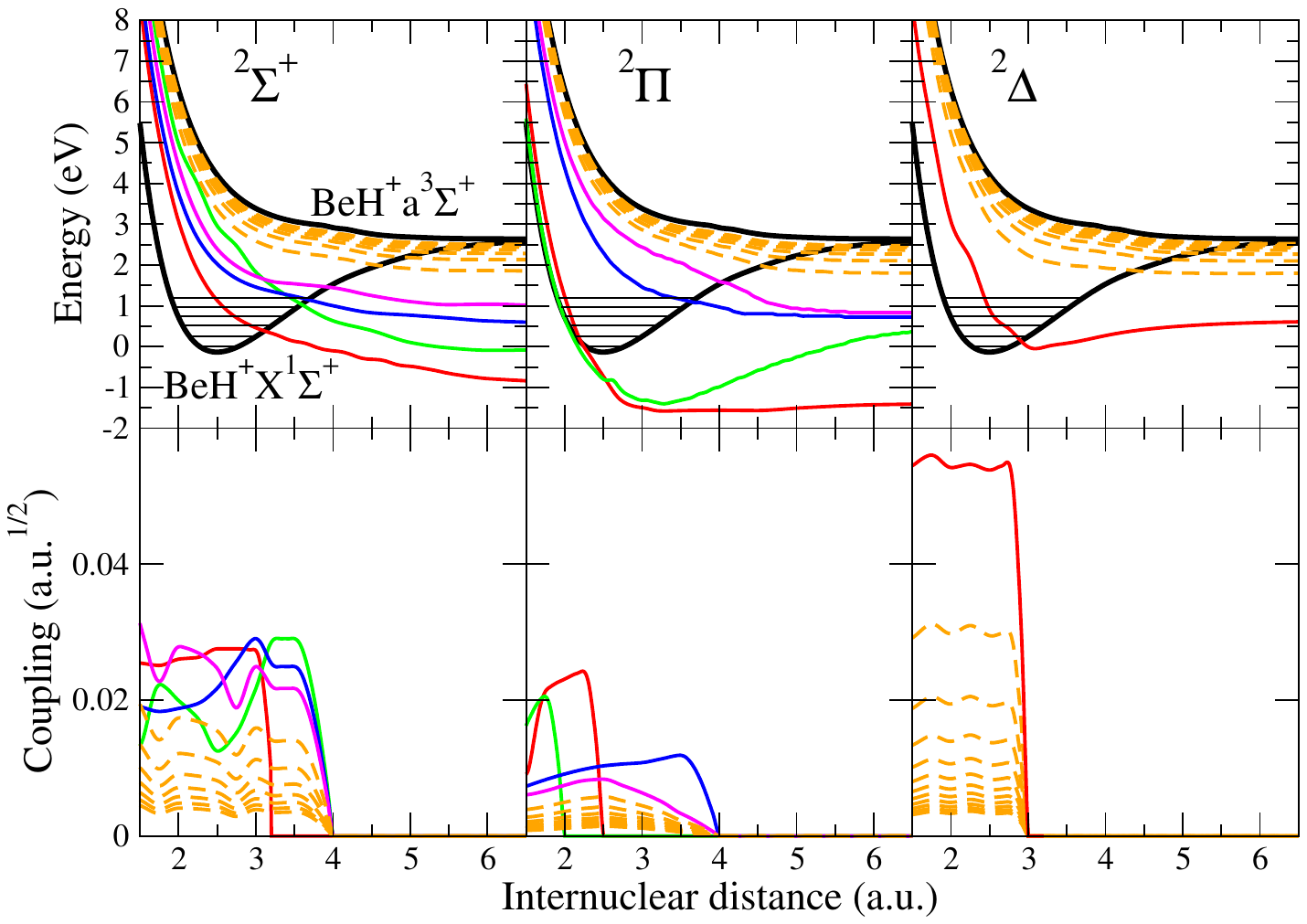}
     \caption{ (Color online) Molecular data relevant for electron-induced reactive processes. Upper row: Potential energy curves of BeH$^{+}$ - thick black lines - and dissociative states of BeH  - continuous coloured lines for {\textit{ab initio}}-obtained and dashed for generated using the scaling laws. 
     Lower row: Electronic couplings between the dissociative states of BeH and the ionization continuum. 
     }
	\label{fig:1} 
\end{figure*} 

In our previous low-energy studies~\cite{Niyonzima2013,Niyonzima, bed, Pop2021}, we used the molecular structure data calculated by Brinne-Roos {\it et al.}~\cite{roos:09} - black, red, green, blue, and magenta curves presented in Fig.~\ref{fig:1} - while the nuclear dynamics calculations were carried out in the framework of the stepwise Multichannel Quantum Defect Theory (MQDT)~\cite{Giusti:80,IFS2000a}.  The incident electron energy range was chosen from  $0.01$ meV to $2.7$ eV, just below the ionic dissociation threshold. In the present paper, we extend our earlier studies to higher collision energies, from $2.7$ up to $12$ eV. This requires the inclusion in the calculation of an increased number of dissociative states, and of the consideration of the DE process, a major competitor to the DR and VT processes. Cross sections and rate coefficients for initial vibrational quantum numbers $v_{i}^{+}=0,1,2$ are provided for each elementary process. 

The paper is organized as follows:  Section~\ref{sec:theory} is dedicated to our theoretical approach for both low- and high-energy collisions, including the building of the required molecular structure data sets.  Section~\ref{sec:results} is devoted to the results of DR, VT, and DE cross sections and rate coefficients for electron collision with the BeH$^+$ ion at low, intermediate, and high energies/temperatures. The paper ends with conclusions.

\section{Theoretical approach}\label{sec:theory}
The stepwise MQDT has been proved to be very suitable in many previous studies on different species, including BeH$^{+}$ and its isotopologues~\cite{Niyonzima, bed, Pop2021}, H$_2^{+}$ and its isotopologues~\cite{MNRAS2016, MNRAS2022}, BF$^{+}$~\cite{mezei2016} for collisions with low as well as high energy~\cite{charkrabati2013} of the incident electron.

The elementary processes (\ref{eq:DR}) -- (\ref{eq:DE}) involve ionization channels - describing the scattering of an electron by the target cation - and dissociation channels - accounting for atom–atom scattering. Below the ionic dissociative threshold, the mixing of these channels results in quantum interference of the direct mechanism - in which the capture takes place into a doubly excited dissociative state of the neutral system - and the indirect one - in which the capture occurs via a Rydberg bound state of the molecule belonging to a closed channel, this state being predissociated by a dissociative one. In both mechanisms, the autoionization - based on the existence of open ionization channels - is in competition with the predissociation and can lead to the excitation or to the de-excitation of the cation.

\subsection{\label{sec:Moldata} Molecular structure data}

In order to describe the electron-molecular cation scattering, a full set of molecular data is required, which includes:  (i) the potential energy curves (PEC) for the cation target BeH$^+$ and the doubly excited dissociative states BeH$^{**}$ of the neutral, (ii) the electronic couplings between  the dissociative states and the ionization continuum (e$^-+$BeH$^+$), and (iii) the quantum defects $\mu$ characterizing the mono-excited BeH$^{*}$ Rydberg states.  

For the neutral system, Roos {\it et al}~\cite{roos:09} have characterized the molecular states for three different symmetries - four states for $^2\Sigma^+$, four states for  $^2\Pi$, and one state for $^2\Delta$ - by combining the multi-reference configuration interaction method with the complex Kohn variational method~\cite{kohn}. In this way, they provided quasidiabatic PECs for the excited electronic states of BeH, including the autoionization widths.

Since the aim of the present paper is to extend the collisional calculations to high energies, we had to provide reasonably well-approximated data for further super-excited states situated above those characterizing the low energies, addressed in our previous studies. Therefore, we have  used the scaling law based on the convergence of the quantum defects - with respect to the first excited state of the cation - with the increase of the principal quantum number of the resonant state. 

For a given symmetry $\Lambda$, labelling  the principal quantum number of the most excited dissociative state computed {\textit {ab initio}} as $n_{h}^{\Lambda}$, and setting $n_{min}^{\Lambda} = n_{h}^{\Lambda} + 1$, we generated the approximate PEC's for the further excited states with principal quantum numbers in the range [$n_{min}^\Lambda$, $n_{max}^\Lambda$] by applying the formula:
\begin{equation}
U^\Lambda_{d,n}(R) = U^{+*}(R) - \frac{Ryd}{[n-{\mu^{\Lambda}_{n_{h}^{\Lambda}}}(R)]^2}\,, 
\end{equation}  
where $n=n_{min}^\Lambda, \ldots, n_{max}^\Lambda$, and $U^{+*}$ represents the PEC of the lowest excited electronic state of  the ion. After checking convergence, we set $n_{max}^\Lambda = n_{min}^\Lambda + 11$ for each symmetry and, consequently, we involve 42 resonant dissociative states in our model. 

As for the  resonance widths corresponding to these generated states, they are given by:
\begin{equation}
\Gamma^\Lambda_{d,n} (R)= \frac{1}{[n-{\mu^{\Lambda}_{n_{h}^{\Lambda}}}(R)]^{3}}\bar{\Gamma}^\Lambda_{d}  (R)
\end{equation}
where $n=n_{min}^\Lambda, \ldots, n_{max}^\Lambda$, and ${\bar\Gamma}^\Lambda_{d}(R)$ is given by :
\begin{equation}
\bar{\Gamma}^\Lambda_{d} (R)= \Gamma_{d,{n_{h}^{\Lambda}}} (R){[{n_{h}^{\Lambda}} -{\mu^{\Lambda}_{n_{h}^{\Lambda}}}(R)]^{3}}\,.
\end{equation}

Once we have made the inventory of the whole set of relevant dissociative states of symmetry $\Lambda$- {\it ab initio} computed, or generated by the Rydberg scaling laws - we compute the electronic coupling between each dissociative continuum associated to the resonant state $j$ (going from 1 to the maximum number of resonant states) and the ionization continuum, by the formula:
\begin{equation}
V^\Lambda_{d_j}(R)
= \sqrt{\frac{\Gamma^\Lambda_{d_j}(R)}{2\pi}}, 
\label{eq:vmat}
\end{equation}
\noindent
where the principal quantum number $n$ is no longer displayed and the $d$ label for 'dissociative' is replaced by the more specific one $d_j$. 

Fig.~\ref{fig:1} presents all the molecular data necessary for the modeling of the quantum dynamics. On the first row, we show the potential energies curves for the \ensuremath{^{2}\Pi}, \ensuremath{^{2}\Sigma^{+}}, and  \ensuremath{^{2}\Delta} electronic symmetries. The black curves are the PECs of the ground states (\ensuremath{X^{1}\Sigma^{+}}) and the lowest repulsive excited state (\ensuremath{a^{3}\Sigma^{+}}) of BeH$^+$. The red, green, blue, and magenta colored curves are the PECs for the lowest resonant curves of BeH obtained by \textit{ab initio} quantum chemistry calculations~\cite{roos:09}. And finally, the dashed orange curves are the new dissociative curves of BeH produced  by the scaling law. On the second row, we present the couplings between the dissociative states of BeH and the ionization continuum for the \ensuremath{^{2}\Pi}, \ensuremath{^{2}\Sigma^{+}}, and  \ensuremath{^{2}\Delta} symmetries. Following the same color code as that of the PECs we can distinguish the couplings  $V^{\Lambda}_{d_j}(R)$, either those obtained by \textit{ab initio} methods ore those produced by the scaling law. All dissociative states and couplings plotted in Fig.~\ref{fig:1} have been included in the calculations.

\subsection{The MQDT method for low-energy collisions}

The low energy region of the electron-molecular cation collisions was explored in our previous papers~\cite{Niyonzima2013, bed, Pop2021}. There, we restricted ourselves to incident electron energies below the ionic dissociation threshold, and we considered the interference of the direct and indirect mechanisms. Here below are the major steps of the modeling of the collision dynamics. 

For each relevant symmetry $\Lambda$ - quantum number describing the projection of the electronic orbital angular momentum on the internuclear axis   - we start by building the interaction matrix  ${\cal V}^{\Lambda}$, whose non-vanishing matrix elements consist in the vibrationally-integrated couplings:
\begin{equation}\label{eq:intmat}
{\cal V}^{\Lambda}_{d_{j},v}(E)= \langle\chi_{d_{j}}(E)|{{V^{\Lambda}_{d_j}}(R)}|\chi_{v}\rangle,
\end{equation}
where $\chi_{d_{j}}$ represents the vibrational wave functions of the dissociative state $d_{j}$, while $\chi_{v}$ is the vibrational wave function for the state of  the ion of  quantum number  $v$ associated with the ionization channel respectively, $E$ is the total energy of the system, and the integration is performed on the geometry (i.e. $R$) range .

In the next stage, we solve the Lippmann-Schwinger matrix equation:
\begin{equation}\label{eq:solveK}
\mathcal{K}^{\Lambda}= \mathcal{V}^{\Lambda} + \mathcal{V}^{\Lambda}{\frac{1}{E-H_{0}}}\mathcal{K}^{\Lambda}.
\end{equation}
in the second order, replacing  $\mathcal{K}^{\Lambda}$ by $\mathcal{V}^{\Lambda}$ in the right-hand side of the preceding equation,  $H_{0}$ being the zero-order Hamiltonian. 

We then build the eigenchannels, using the diagonalized version of the reaction matrix $\mathcal{K}^{\Lambda}$, and we perform eventually a frame transformation~\cite{IFS2000a} from the internal/reaction zone to the external/asymptotic region:
\begin{equation}\label{eq:solve1}
\mathcal{C}^{\Lambda}_{v,{ \alpha}}=\\
\sum_{v'}U^{\Lambda}_{v',\alpha}\langle\chi_{v}(R)|\cos(\pi \mu^{\Lambda}(R)+ \eta^{\Lambda}_{\alpha})|\chi_{v'}(R)\rangle,
\end{equation}
\begin{equation}\label{eq:solve2}
\mathcal{C}^{\Lambda}_{d, \alpha}=U^{\Lambda}_{d\alpha} \cos\eta^{\Lambda}_{\alpha},
\end{equation}
and similar $\mathcal{S}$  matrix elements obtained by replacing cosine with sine in Eqs.~(\ref{eq:solve1}) and~(\ref{eq:solve2}). Here $\alpha$ designates the eigenchannels defined through the diagonalization of the  $\mathcal{K} ^{\Lambda}$-matrix,$U_{d\alpha}^{\Lambda}$ and $U_{lv,\alpha}^{\Lambda}$ are elements of the corresponding eigenvectors, and $-\frac{1}{\pi}\tan{\eta}^{\Lambda}_{\alpha}$ are the eigenvalues of ${\mathcal{K}} ^{\Lambda}$.

The projection matrices $\mathcal{C} ^{\Lambda}$ and $\mathcal{S} ^{\Lambda}$ are used to build the generalized scattering matrix $X ^{\Lambda}$ relying on both energetically open  $o$ and closed $c$ channels: 
\begin{equation}\label{eq:Xmatrix}
X^{\Lambda}=\frac{\mathcal{C}^{\Lambda}+i\mathcal{S}^{\Lambda}}{\mathcal{C}^{\Lambda}-i\mathcal{S}^{\Lambda}},
\qquad
X^{\Lambda}= \left(\begin{array}{cc} X_{oo}^{\Lambda} & X_{oc}^{\Lambda}\\
                   X_{co}^{\Lambda} & X_{cc}^{\Lambda} \end{array}\right).
\end{equation}

 Applying Seaton's method on the elimination of the closed channels~\cite{Seaton1983} we finally obtain the scattering matrix:
 \begin{equation}\label{eq:solve3}
S^{\Lambda}=X_{oo}^{\Lambda}-X_{oc}^{\Lambda}\frac{1}{X_{cc}^{\Lambda}-\exp(-2i\pi 
{\bf \beta})}X_{co}^{\Lambda},
\end{equation}
where the diagonal matrix $\exp(-2i\pi {\bf \beta})$ contains the effective quantum numbers $\beta_v$ corresponding to the vibrational ($v$) thresholds of the closed ionization channels at the given total energy of the system. 

Once the scattering matrix is known, the DR and VT cross sections for a particular angular momentum quantum number $ \Lambda$ are given by
\begin{equation}\label{eq:solv5}
\sigma_{diss \leftarrow v_{i}}^{\Lambda}=\frac{\pi}{4\varepsilon} \rho^{\Lambda}\sum_{l,j}|S_{d_{j},lv_{i}}|^2,
\end{equation}
\begin{equation}\label{eq:solv6}
\sigma_{v_{f} \leftarrow v_{i}}^{\Lambda}=\frac{\pi}{4\varepsilon} \rho^{\Lambda}\sum_{l,l'}|S_{l' v_{f},lv_{i}}-\delta_{l^{'}l}\delta_{v_{f}v_{i}}|^2 .
\end{equation}
 
Summing over all possible molecular symmetries, the global DR and VT cross sections are obtained:
\begin{equation}\label{eqDR}
\sigma_{diss \leftarrow v_{i}}=\sum_{\Lambda} \sigma_{diss \leftarrow v_{i}}^{\Lambda},
\end{equation}
\begin{equation}\label{eqVE_VdE}
\sigma_{v_{f} \leftarrow v_{i}}=\sum_{\Lambda} \sigma_{v_{f}\leftarrow v_{i}}^{\Lambda},
\end{equation}
where $\varepsilon$ is the incident electron energy, $\rho^{\Lambda}$ is the ratio between the state multiplicities of the neutral system and of the ion.

\subsection{The MQDT method for high-energy collisions}

At energies higher than the ionic dissociation threshold, we have to take into account the autoionization resulting into states from the continuum part of the vibrational spectrum of the cation ground state - which we will call from now on 'core 1' - leading to dissociative excitation (DE), Eq.~(\ref{eq:DE}). When this process is included in our approach, the couplings between a given dissociation channel $d_j$ and an ionization channel $v$ (Eq.~(\ref{eq:intmat})) is extended to the continuum part of the vibrational spectrum. This is achieved by discretizing the continuum by using the Fourier-grid method~\cite{Fourier}, which relies on the grid-points which span the range of the internuclear distance relevant for the present work. The grid method gives at the same time, in one calculation, the full vibrational ladder. In order to achieve convergence of the cross section for the high-energy region, we allowed for 180 quasi-continuum vibrational levels in the continuum.

This procedure has been applied for the ground electronic state of BeH$^+$ - the so-called 'core 1' - but also for the lowest excited electronic state, 'core 2', of $^3\Sigma^+$ symmetry,  having a repulsive PEC with the same dissociation limit as core 1, cf. Fig.~\ref{fig:1}. Every ionization channel $v$ built on core 1, effective below or above the dissociation threshold, is coupled to further ionization channels built on core 2, labeled  by $w$, situated entirely in the continuum:
\begin{equation}
	{\mathcal V}_{w v}^{\Lambda}  = \langle \chi_{w}^\Lambda|
	\tilde{\it{V}}^{\Lambda}(R)|\chi_{v}^\Lambda \rangle
. \label{eq:Vwv}
\end{equation}
\noindent
where $\tilde{\it{V}}^{\Lambda}(R)$ - assumed to be energy independent - is the electronic coupling between the two ionization continua. 

Taking all the above into account, the extended $\mathcal{K}$-matrix has the following form (the index $\Lambda$ being omitted):
\begin{equation}\label{Kmat2}
\mathcal{K} = \left( \begin{array}{ccc}
\mathcal{K}_{\bar d\bar d} & \mathcal{K}_{{\bar d}\bar v} & \mathcal{K}_{\bar d \bar w}\\
\mathcal{K}_{\bar v \bar d} & \mathcal{K}_{\bar v \bar v} & \mathcal{K}_{\bar v \bar w}\\
\mathcal{K}_{\bar w \bar d} & \mathcal{K}_{\bar w \bar v} & \mathcal{K}_{\bar w \bar w}\\
\end{array} \right).
\end{equation}
where the collective indices $\bar d$, $\bar v$, $\bar w$ in Eq.~(\ref{Kmat2}) span the ensembles of all available dissociation and ionization channels, with the latter ones built on cores 1 and  2 respectively.  The first two rows and columns of the K-matrix are responsible for the DR at low energy, while the third row and the third column complete the $\mathcal{K}$-matrix for the correct description of the DR at high energy, including the DE.

Similarly to the low-energy case, the Lippmann-Schwinger equation for the $\mathcal{K}$-matrix can be solved exactly in second order assuming the interactions ${V_{d_j}}^{(e)\Lambda}$ and $\tilde{\it{V}}^{(e)\Lambda}(R)$ independent on the energy of the incident electron.

\noindent For a given $\Lambda$ symmetry and for all $v$ and $w$ bound and continuum vibrational levels, all the interaction matrix elements except those given by Eqs. (\ref{eq:intmat}) and (\ref{eq:Vwv}), i.e. $V^{\Lambda}_{d_i d_j}$, $V^{\Lambda}_{d_i w}$, $V^{\Lambda}_{v v'}$ and $V^{\Lambda}_{w w'}$, vanish,, since in the quasidiabatic representation chosen here, they concern pairs of channels  associated to the same ionic core~\cite{0022-3700-4-8-008}. Taking into account the non-vanishing matrix elements related to the available dissociation and ionization channels, the second order $\mathcal{K}$-matrix writes~\cite{Kalyan2012},
\begin{equation}\label{Kmat3}
\mathcal{K} = \left( \begin{array}{ccc}
\mathit{0} & \mathcal{V}_{\bar d \bar v} & \mathit{0}\\
\mathcal{V}_{\bar v \bar d} & \mathcal{K}_{\bar v \bar v} & \mathcal{V}_{\bar v \bar w}\\
\mathit{0}& \mathcal{V}_{\bar w \bar v} & \mathit{0} \\
\end{array} \right)
\end{equation}
where $\it{O}$ represent  null matrices.

The inclusion of dissociative excitation induced by vibronic continuum levels of the two cores increases not only the dimensions of the interaction matrix $\mathcal{V}$~(Eq.~\ref{eq:Vwv}) and the reaction $\mathcal{K}$-matrix (\ref{Kmat2}), but also those of the frame transformation $\mathcal{C}$ and $\mathcal{S}$ matrices. Once we have the above-mentioned matrices, the scattering matrix is calculated following the steps presented in the previous subsection, according to Eqs.~(\ref{eq:Xmatrix}) and~(\ref{eq:solve3}) respectively. The DE1 and DE2 processes are eventually treated as vibrational excitations from an initial vibrational state $v_i$, to the discretized ionization continua of the two cores. The cross sections for these processes for each molecular symmetry are given by:
\begin{equation}\label{DE1}
	\sigma_{DE1,\; v_i}^{\Lambda}=  \frac{\pi}{4\varepsilon}
	\rho^{\Lambda} \sum_{v_h <v < v_{max}(\varepsilon)} |S_{v, v_i}|^2,
\end{equation}
\begin{equation}
\label{DE2}
	\sigma_{DE2,\; v_i}^{\Lambda}=  \frac{\pi}{4\varepsilon}
	\rho^{\Lambda} \sum_{w < w_{max}(\varepsilon)} |S_{w, v_i}|^2,
\end{equation}
\noindent
where $v_h$ is the highest \textit{bound} vibrational level built on core 1, while $v_{max}(\varepsilon)$ and  $w_{max}(\varepsilon)$ are the highest \textit{quasi-continuum} vibrational levels situated below the current total energy $E=E_{v_{i}}+\varepsilon$, corresponding to core 1 and core 2 respectively.
\begin{figure}[t]
     \centering\includegraphics[width=\linewidth]{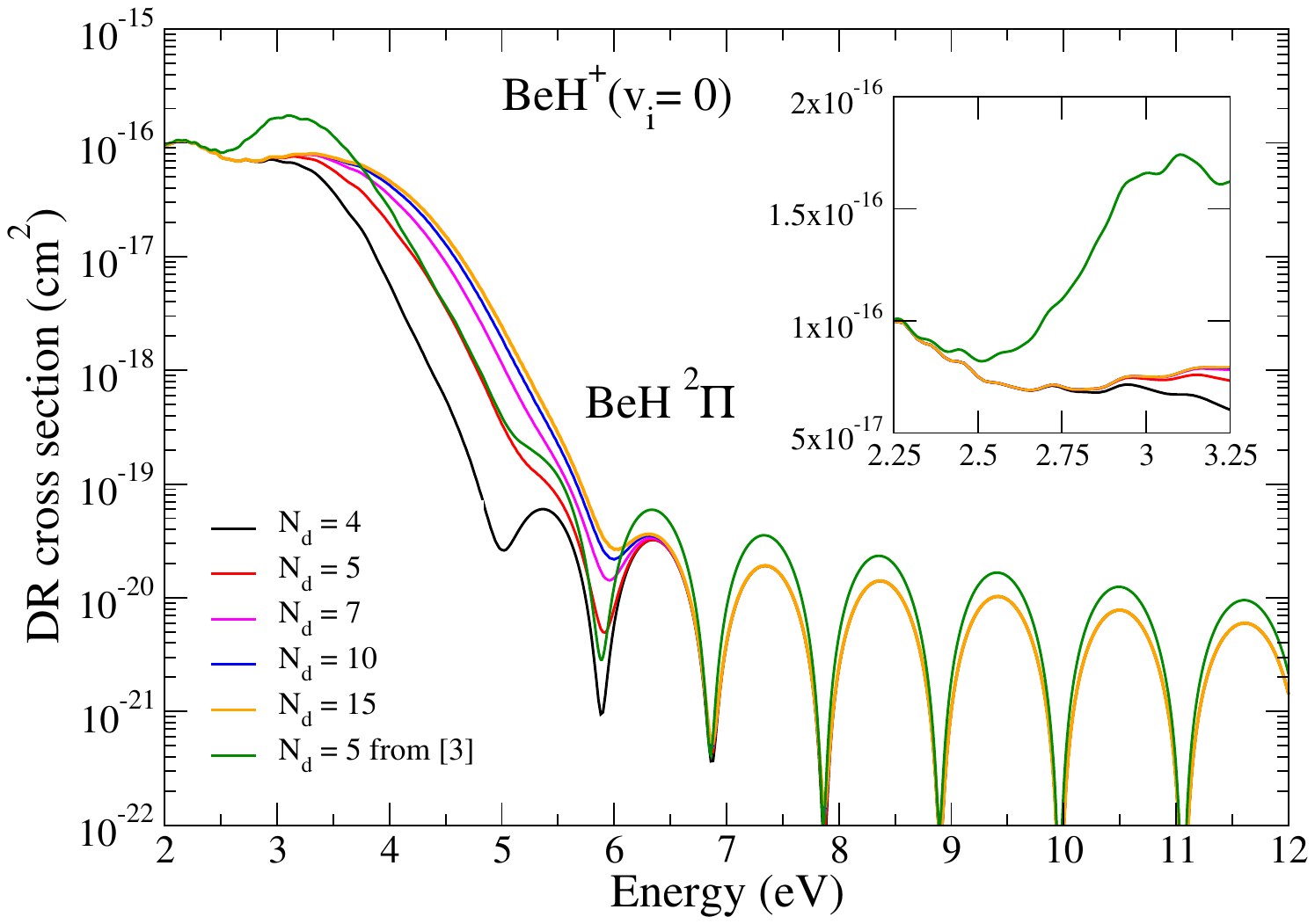} 
\caption{ (Color online) 
Convergence of the DR cross section with respect to the number of dissociative states $N_{d}$, for the case of vibrationaly relaxed ($v_i=0$) BeH$^{+}$ in its ground electronic state $X\ensuremath{^{1}\Sigma^{+}}$, by considering only the $^2\Pi$ symmetry of the neutral. 
}
\label{fig:2}
\end{figure}
 \begin{figure}[t]
    \centering\includegraphics[width=\linewidth]{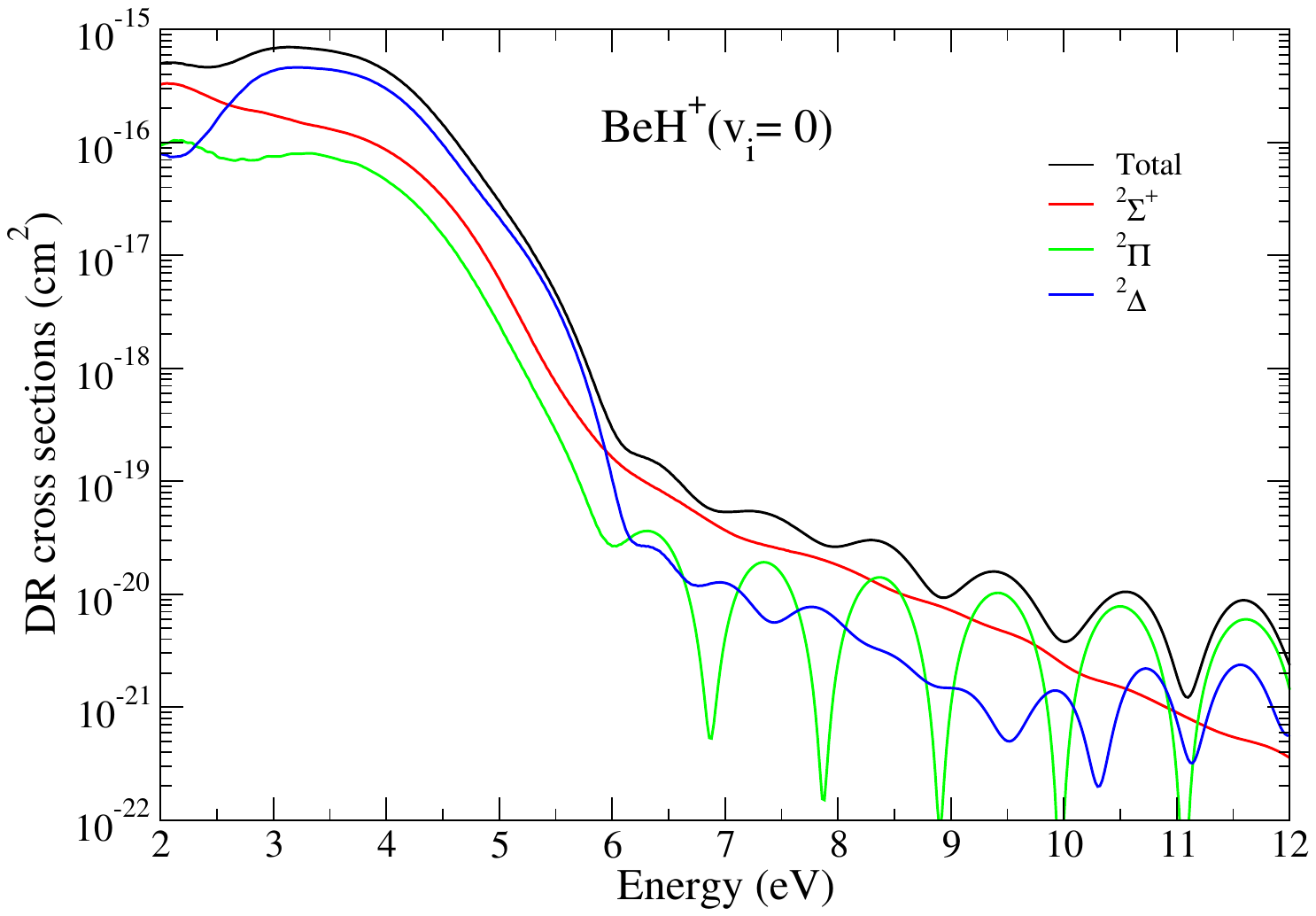}     
     \caption{ (Color online) Direct  DR cross sections of BeH$^{+}$ in its ground vibrational state: Contribution of each symmetry.
     }
\label{fig:3}
\end{figure}

The total DE cross section is obtained by the sum of DE1 and DE2 for each symmetry,  further summed up over all relevant molecular symmetries:
\begin{equation}\label{eqDE}
	\sigma_{DE,v_i} = \sum_{ \Lambda}
\sigma_{DE, v_i}^{\Lambda} = \sum_{ \Lambda} \left( \sigma_{DE1,\; v_i}^{\Lambda} + \sigma_{DE2,v_i}^{\Lambda}\right)
.
\end{equation}
\noindent
And finally, the cross sections for the DR and VT processes are those given in Eqs.~(\ref{eq:solv5})--(\ref{eqVE_VdE}), since they do not change with the inclusion of the discretised vibrational continua, in spite of the fact that the vibrational continua and  the generated high-lying dissociative states are included in the structure of the scattering matrix.
 
\begin{figure}[t]
    \centering
    \includegraphics[width=\linewidth]{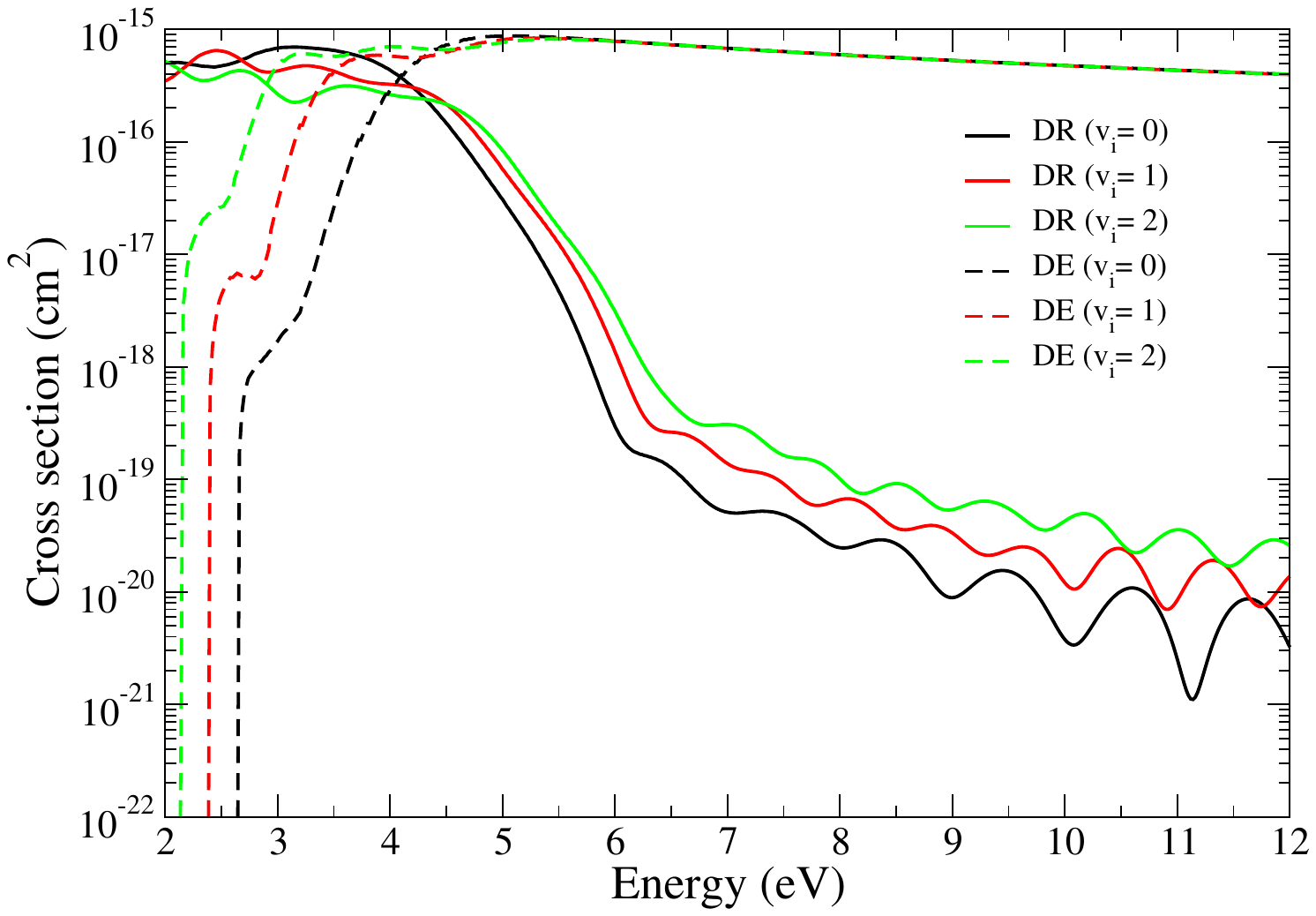}     
     \caption{ (Color online) Direct  DR (continuous lines) and DE (dashed lines) cross sections of BeH$^{+}$ in its electronic ground state for initial vibrational levels $v_{i}=0,1$ and $2$.}
\label{fig:4}
\end{figure}
 \begin{figure}[t]
     \centering\includegraphics[width=\linewidth]{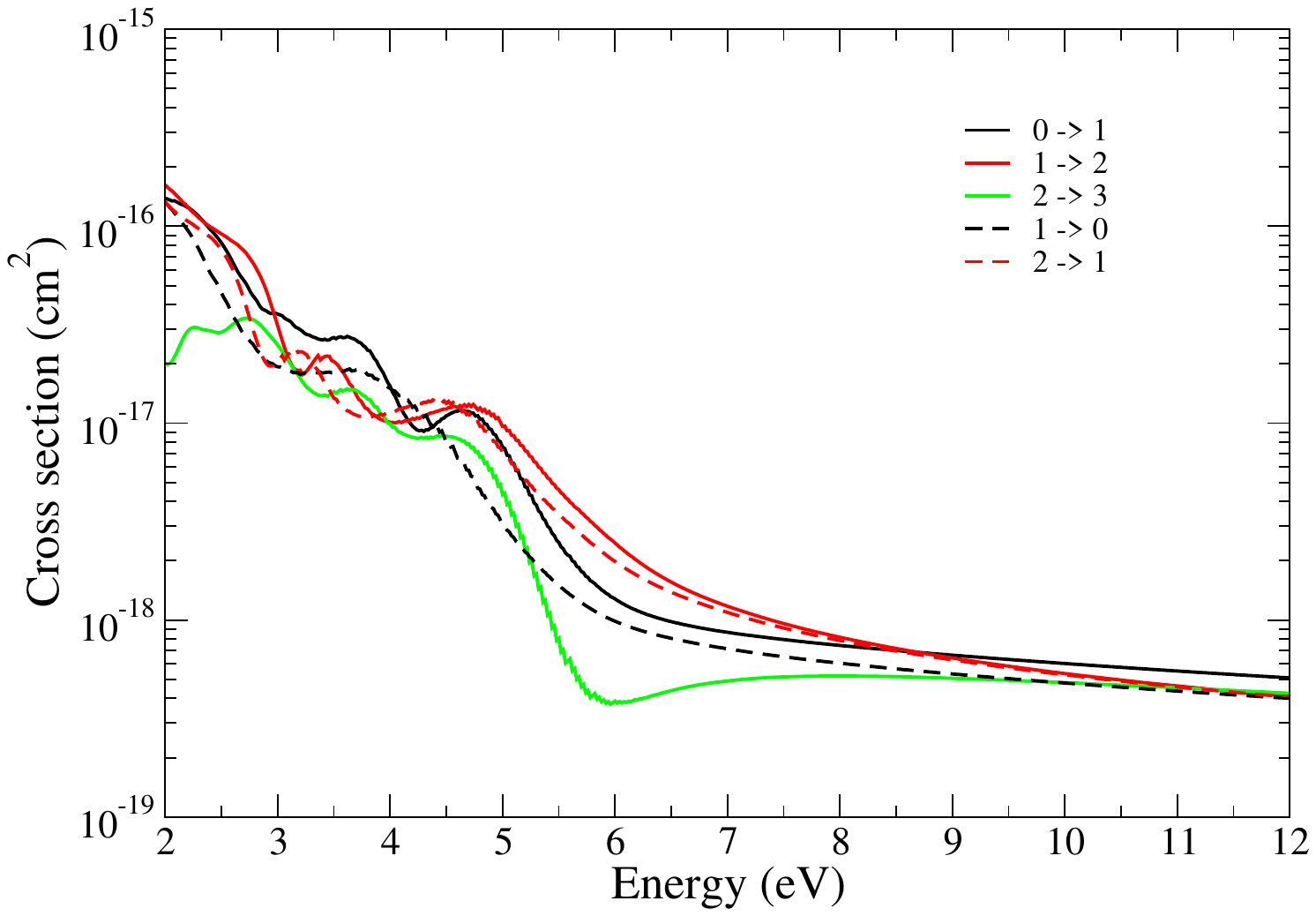}
\caption{ (Color online)  VE (continuous lines) and VdE(dashed lines) cross sections for BeH$^{+}$ $(v_{i}=0, 1, 2)$ for $\Delta v= v_{f}-v_{i}= \pm 1$ transitions.}
\label{fig:5}
\end{figure}

\section{Results and discussions}\label{sec:results}

\begin{figure*}[t]
    \centering
    \includegraphics*[width=0.75\linewidth]{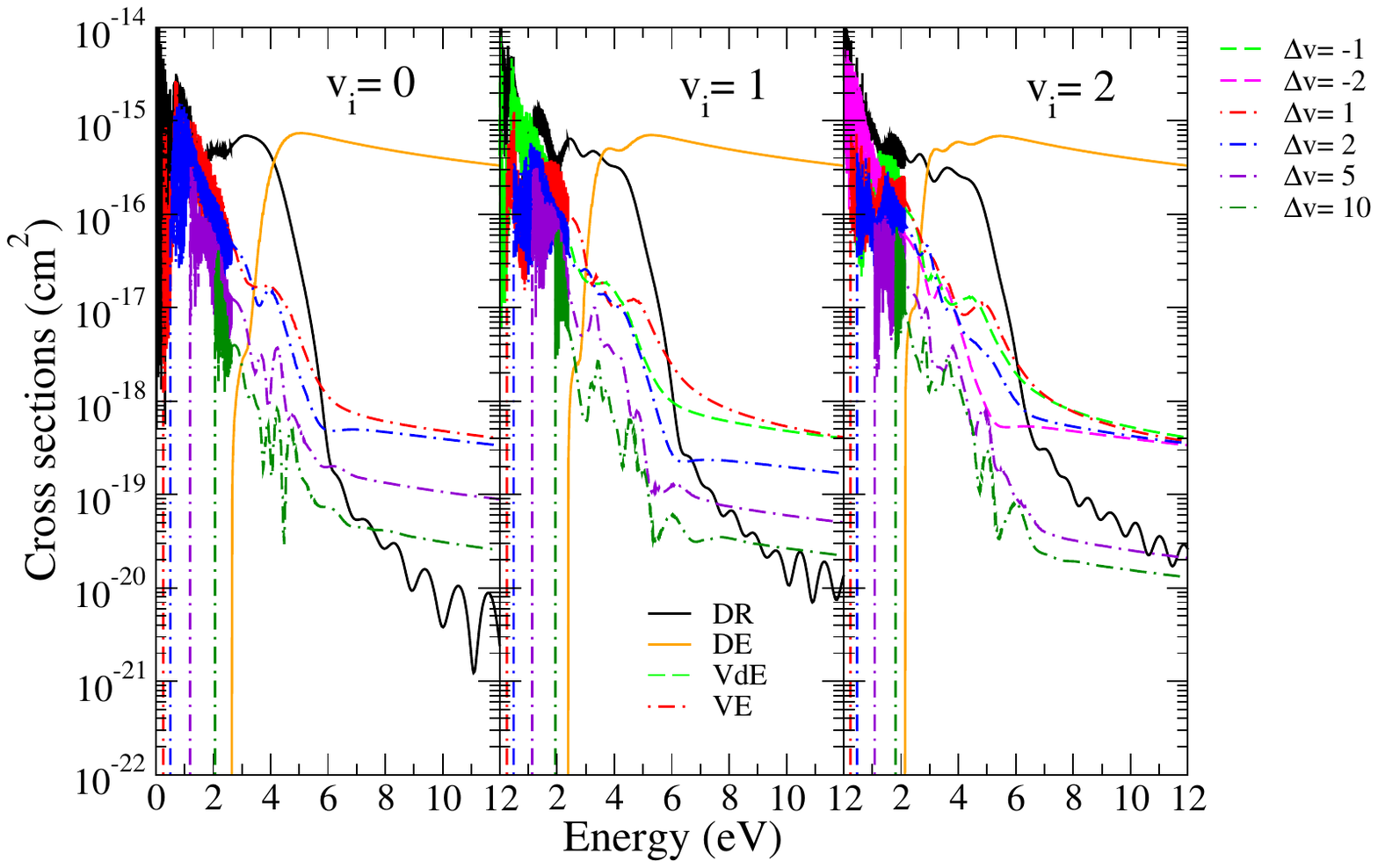}     
     \caption{ (Color online) DR, DE, VdE, VE cross sections for BeH$^{+}$ in its ground electronic states for the lowest three vibrational levels of the target ($v_{i}=0, 1, 2$). $\Delta v=v_{f}-v_{i}$}
\label{fig:6}
\end{figure*}

\begin{figure*}[t]
\centering
\includegraphics[width=0.75\linewidth]{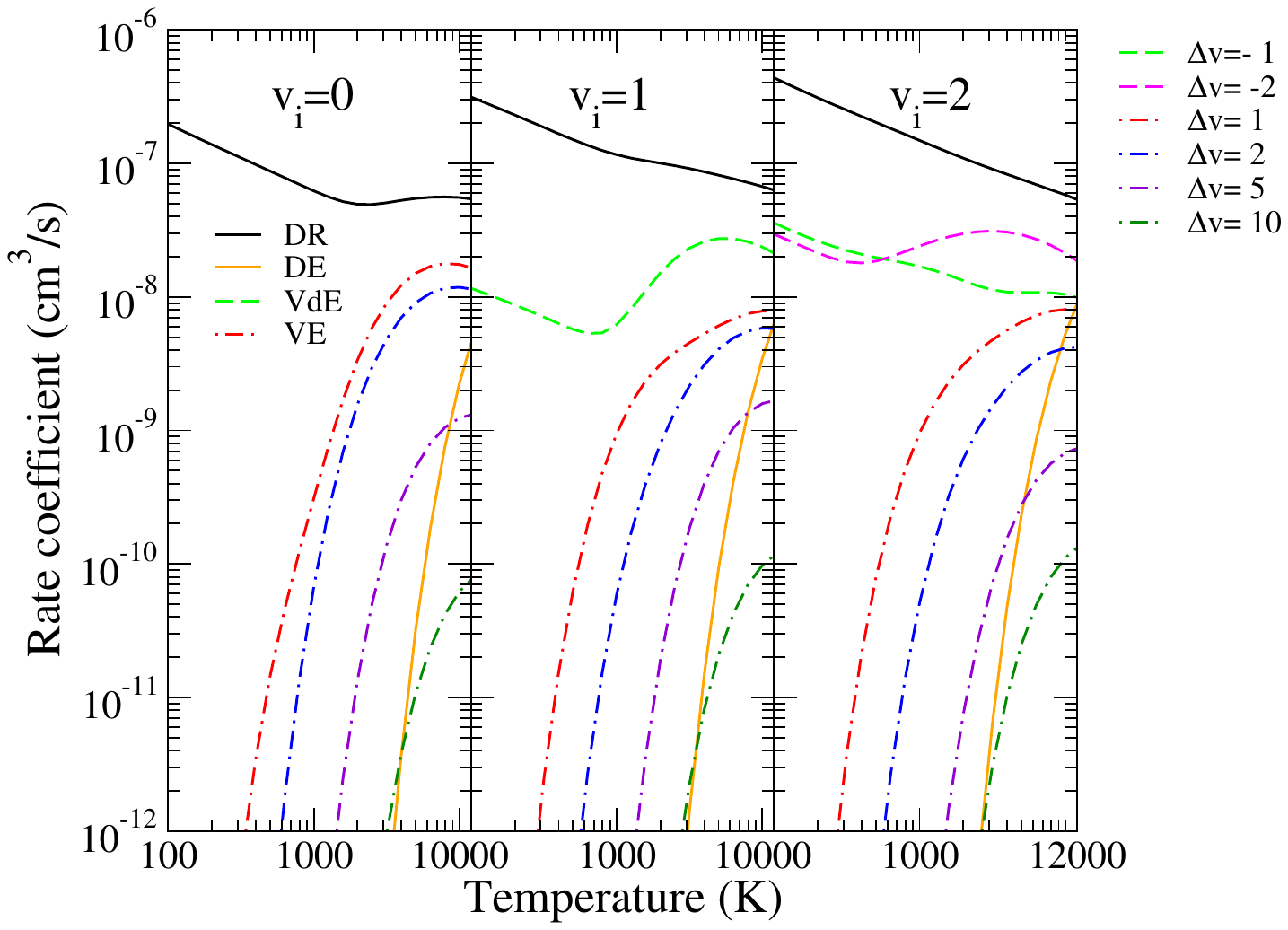}
        \centering\caption{ (Color online)  DR, DE, VdE, VE rate coefficients of BeH$^{+}$ in its ground electronic states for the lowest three vibrational levels of the target ($v_{i}=0, 1, 2$). $\Delta v=v_{f}-v_{i}$}
\label{fig:7}
\end{figure*}

Applying the stepwise MQDT method outlined in the previous section we have calculated the dissociative recombination (Eq.~(\ref{eq:DR})),  vibrational transition (excitation and de-excitation) (Eq.~(\ref{eq:VT})) and dissociative excitation (Eq.~(\ref{eq:DE})) cross sections of BeH$^+$ for its lowest three ($v_i = 0-2$) vibrational levels of its ground electronic state. The electron impact collision energies range between 0.01 meV and 12 eV, with a step of $0.02$meV for energies below the ionic dissociation limit and with a step of 1 meV for energies above the limit. These cross sections are then used to obtain thermal rate coefficients for electron temperatures ranging between 10 and 12000 K. The calculated DR, DE, VE and VdE cross sections are given in Figs.~\ref{fig:2} to~\ref{fig:6}, while the evaluated rate coefficients in Figs~\ref{fig:7} to~\ref{fig:9}.

\subsection{Cross sections} 
At energies below the dissociation threshold of the target, the cross section for dissociative recombination and related processes are as the result of the quantum interference of the direct and indirect mechanisms. The direct mechanism is responsible for the smooth, bulk part of the cross section, while the indirect mechanism is responsible for the infinite number of resonances in the cross section. This behavior can be most easily understood according to Eq.~(\ref{eq:solve3}), where the first term is provided by the direct mechanism, while the second term containing closed ionisation channels is the indirect one. Moreover, the direct term is directly connected to the Franck-Condon overlap between the vibrational wave function of the dissociative state, the wave function of the target, and the vibronic interaction among the dissociation and ionization channels, confirm Eq.~(\ref{eq:intmat}). The indirect mechanism loses its importance for higher collision energies due to the lack of bound vibrational levels of the mono-excited Rydberg states. Thus one can conclude that for collision energies higher than the ionic dissociation threshold, the total DR cross section reduces to the direct cross section only. This is visible for example in Fig.~\ref{fig:6}, where the Rydberg resonances disappear from the cross section as soon as we achieve the ionic dissociation threshold. In what follows we will discuss the dependence of our results on the number of dissociation states or different molecular symmetries included in the calculation for collisional energies ranging from 2 eV up to 12 eV.

The  convergence of the DR cross sections obtained for the vibrationally relaxed BeH$^+$ target with the number of generated dissociative states is presented in Fig.~\ref{fig:2}. 
DR cross sections - direct mechanism only calculated for the most relevant \ensuremath{^{2}\Pi} molecular symmetry - are given as a function of the collision energy - ranging from 2 eV up to 12 eV - in a log-linear scale, for different numbers of dissociative states. The black curve stands for the DR cross section obtained with the original {\it ab initio} molecular states of Brinne-Roos {\it et al}~\cite{roos:09} ($N_d=4$). In contrast, the red, magenta, blue, and orange curves stand for the present cases having $N_d=5,7,10$ and $15$ generated dissociative states. These results are compared with the cross sections obtained in a previous study~\cite{Niyonzima2013}, where a global fifth dissociative state was generated starting from the {\it ab initio} states, labeled as $N_d=5$ and shown in dark green in Fig.~\ref{fig:2}, 

We obtained very good agreement with the previous low-energy cross sections up to 2.3 eV collision energies according to the inset of Fig.~\ref{fig:2}. In contrast, starting from the ionic dissociation threshold we find differences up to a factor of two. The explanation for this lies in the two different ways of generating the $N_d=5$ and higher dissociative states. The shape of the cross sections is due to the Franck-Condon overlap among the dissociation, and target vibrational wave functions and vibronic interactions. The fall of the cross sections around 4 eV is due to the small vibronic interaction (see Eq.~(\ref{eq:intmat})), while the dissociative wave function is responsible for the oscillations above 6 eV.

We have performed a convergence study by continuously increasing the number of generated dissociative states, by imposing an overall change in the DR cross section of less than $5\%$. This criterion led us to 11 generated states for each of the  molecular symmetries (\ensuremath{^{2}\Sigma^{+}}, \ensuremath{^{2}\Pi}, and \ensuremath{^{2}\Delta}) relevant in the present calculations. 
 
 In Fig.~\ref{fig:3} we present the direct DR cross sections of BeH$^{+}$ in its ground vibrational level, illustrating the contribution of each molecular symmetry \ensuremath{^{2}\Sigma^{+}}, \ensuremath{^{2}\Pi}, \ensuremath{^{2}\Delta} represented by solid red, green and blue curves, to the global - black curve in the same figure - cross section obtained as the sum over the three symmetries. Due to the more favorable crossing of the dissociative states with the ground state of the target, for collision energies below the ionic dissociation threshold, the $^2\Sigma^+$ symmetry gives the largest contribution to the global DR cross section, followed by the $^2\Pi$ and $^2\Delta$ symmetries. Between the ion dissociation limit and 6 eV, the leading symmetry becomes the $^2\Delta$. For still higher energies, the leading symmetry is either the $^2\Pi$ or the $^2\Sigma^+$ one.

For energies above the ionic dissociation threshold, the vibrational continuum is discretized, allowing the consideration of the dissociative excitation (Eq.~(\ref{eq:DE})). Fig.~\ref{fig:4} shows the total DR (black, red, and green curves) and DE (blue, magenta, and cyan curves) cross sections obtained for the three lowest ($v_{i}=0,1$ and $2$) vibrational levels of the target. One can notice that the behavior of the two dissociative processes is completely different. While the DR cross section following a plateau starts to decrease for collision energies approximately 1 eV above the ionic threshold and vanishes essentially above 6 eV, the DE cross section shows a strong threshold effect at the ionic dissociation energy and starts to increase, arriving at the magnitude of DR around 4 eV. Up to this energy point the total dissociation flux coming from the reaction zone is shared between the DR and DE processes, while starting from 6 eV that almost totally goes into DE. Moreover, starting from this energy, the DE cross sections show a smoothly decreasing behaviour.

Besides the dissociation channels given in Eqs.~(\ref{eq:DR}) and~(\ref{eq:DE}) we have studied their competitive vibrational transitions too (Eq. (\ref{eq:VT})). Fig.~\ref{fig:5} displays the $\Delta v=\pm1$ vibrational transitions (excitations and de-excitations) cross sections from the lowest three vibrational levels of BeH$^{+}$ for collision energies ranging from 2 eV to 12 eV. The black, red and green curves stand for the $0\rightarrow 1$, $1\rightarrow 2$ and $2\rightarrow 3$ vibrational excitations, the magenta and cyan curves represent the $1\rightarrow 0$ and $2\rightarrow 1$ vibrational de-excitations respectively. The cross section show slight oscillations due to the Franck-Condon overlap among the bound vibrational wave functions of the target, and tends towards zero at higher energy.
 
Finally, in Fig.~\ref{fig:6} we illustrate the relative importance of all these processes for the lowest three initial vibrational levels of the target by representing their cross sections for collision energies ranging between $0.01$ meV to 12 eV. The excitation and de-excitation cross sections have been calculated for all possible final states, in the figure we present only a few cases belonging to $\Delta v^+=1,2,5$ and $10$ excitations (dashed-dotted red, blue, violet, and dark green curves) and $\Delta v=-1$ and $-2$ de-excitations (dashed light green and magenta curves). One can see that for low collision energies, below the ionic dissociation threshold, the DR (black curves), VE and VdE cross sections are characterised by rich resonant structures due to the temporary capture into Rydberg states, but as we already discussed in our previous studies~\cite{Niyonzima2013,Niyonzima}, in overall they have a weak contribution to the total cross section. The oscillations in the VT cross sections observed at collision energies above the ionic dissociation threshold are due to the Franck-Condon overlap among bound vibrational wave functions. Similarly to DE (orange curves), the vibrational excitation processes show strong thresholds at their respective excitation energies. The DR cross section shows a strong dependence on the initial vibrational level of the cation, due to the difference in vibrational wave functions of the initial state and due to the more favorable crossings of the dissociative state for $^2\Delta$ symmetry with the higher lying vibrational levels of the cation. In the low collision energy range, the DR is in competition with the VT processes, while at energies above the ionic dissociation threshold, only the two dissociation channels compete. Starting from the ionic dissociation threshold the DR process continues to be the dominant process, while following its excitation threshold the DE is becoming more and more important and around 4 eV will take over the lead. 

\subsection{Rate coefficients}

 \begin{figure}[t]
\centering\includegraphics[width=\linewidth]{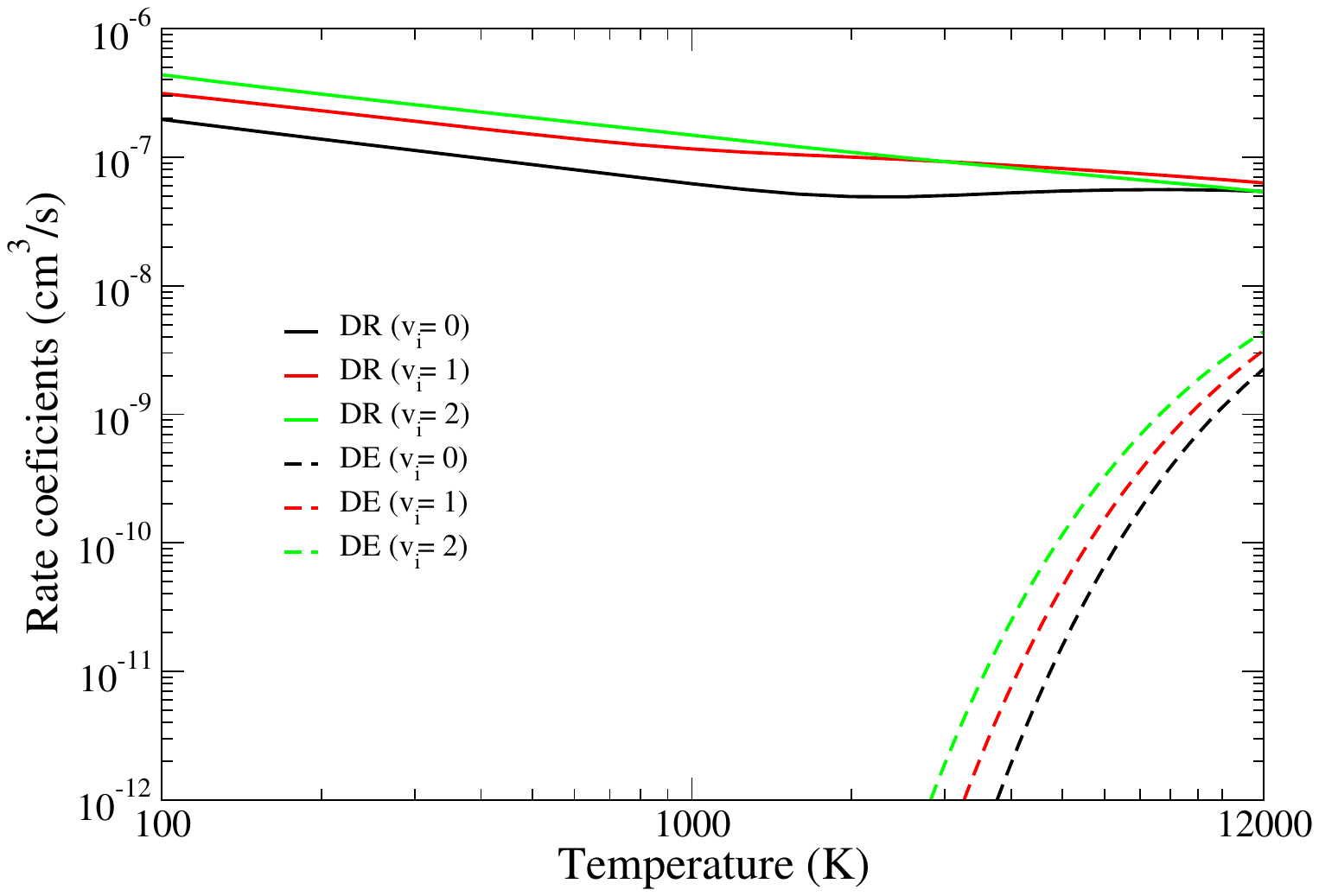}
          \centering\caption{ (Color online)  DR (continuous lines) and DE (dashed lines) rate coefficients of BeH$^{+}$ for $v_{i}=0,1,2.$ calculated using \ref{rate}from the cross sections shown in the Fig.\ref{fig:4} }
\label{fig:8}
\end{figure}
\begin{figure}[t]
\centering\includegraphics[width=\linewidth]{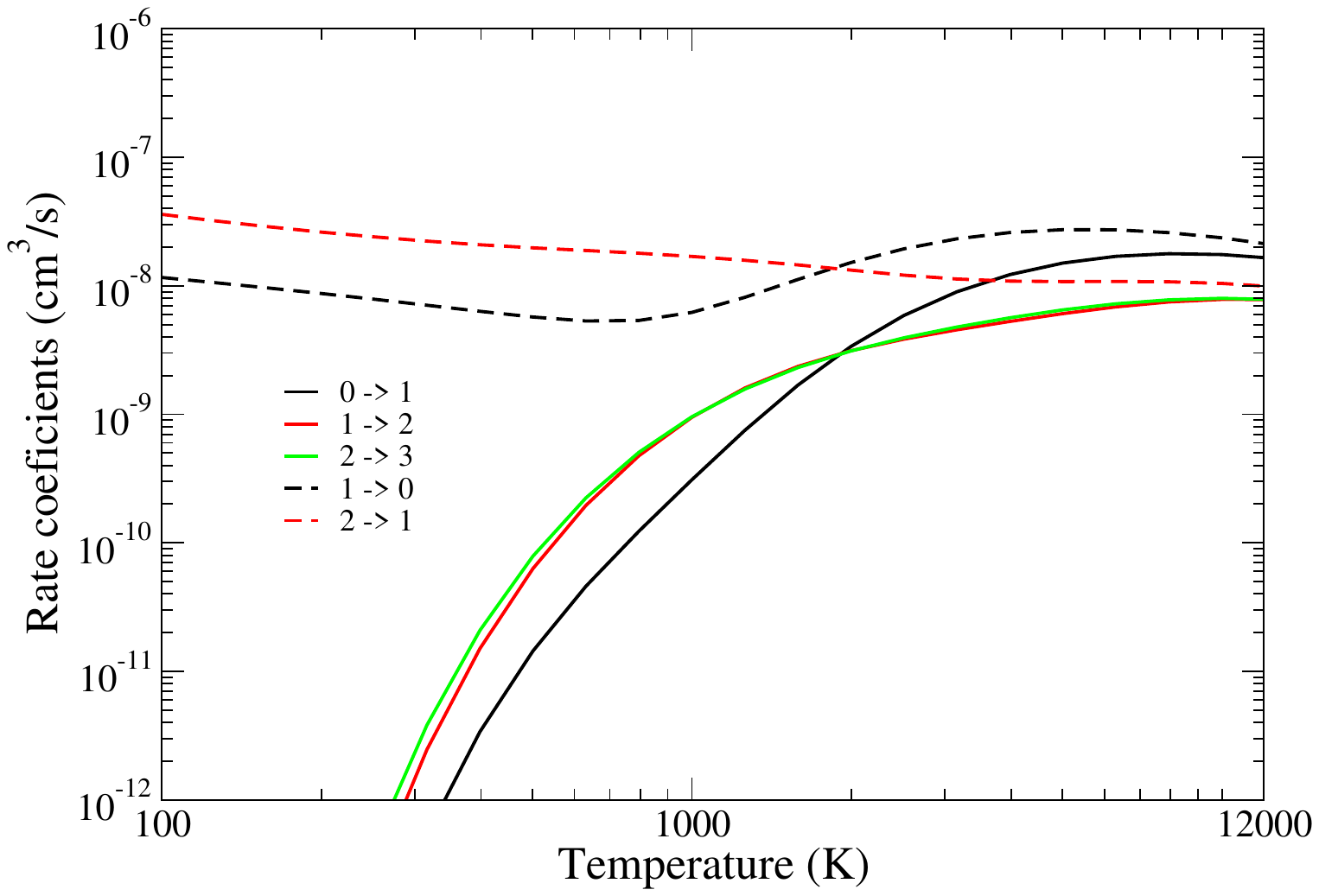}
          \centering\caption{ (Color online) VE(continuous lines) and VdE(dashed lines) cross sections of BeH$^{+}$ $(v_{i}=0, 1, 2)$ calculated using \ref{rate}from the cross sections shown in the Fig.\ref{fig:5}.}
\label{fig:9}
\end{figure}
By convoluting the cross sections with the Maxwell isotropic velocity distribution function of the free electrons~\cite{Mathur1971} we obtain the thermal or plasma rate coefficients:
\begin{equation}\label{rate}
\alpha(T)=\frac{2}{kT}\sqrt{\frac{2}{\pi mkT}}\int_{0}^{\infty}\sigma(\varepsilon)\varepsilon\exp(-\varepsilon/kT)d\varepsilon,
\end{equation}
 where $T$ stands for the electron temperature, $\sigma(\varepsilon)$ is the cross sections for the DR, VE, VdE and DE processes given by equations~(\ref{eqDR}),~(\ref{eqVE_VdE}) or~(\ref{eqDE}), $m$ is the mass of the electron and $k$ is the Boltzmann constant.

The rate coefficients corresponding to the cross sections presented in Fig.~\ref{fig:6} are shown in Fig.~\ref{fig:7}, where the same color code was used. The rate coefficients are given for all~(\ref{eq:DR})--(\ref{eq:DE}) competitive processes having the BeH$^{+}$ ion target in its lowest three vibrational levels. All DR, DE, VdE rate coefficients, and only a few pertinent VE cases ($\Delta v =1,2,5$ and $10$) are shown. The electron temperature varies from 100 K to 12000 K, and we use the same color code as given in Fig.~\ref{fig:6}.

The thermal rate coefficients in contrast to the cross sections have smoother energy/temperature dependence, giving a more adequate description of the importance of each process. According to fFig.~\ref{fig:7}, the DR is the most important reactive process for the presented temperature range, followed by the vibrational de-excitations. The rates for the excitations (VE and DE processes) are reaching the magnitude of the DR/VdE rates only at higher temperatures, close to or above 12000 K. Their very sharp thresholds visible in the cross section are turned into monotonically increasing functions. 

Moreover, Fig.~\ref{fig:7} shows the dependence of the rate coefficients on the initial vibrational levels of the target, presented here for $v_{i}=0,1,2$.
Depending on the vibrational quantum number of the initial and final states of the target, the rates show various temperature dependencies, from the smoothly decreasing behavior to more general functions showing at least one minimum/maximum. For DR, DE, and VdE, the magnitude of rate coefficients increases with the initial vibrational quantum numbers, while for VE one can observe the converse. 

It is the very different dependence of the DR, DE, VE, and VdE processes on the
initial and final channels that leads to very different energy/temperature dependence of the cross sections/rate coefficients. The most complex initial/final channel dependence can be observed for the DR and DE (see Eq.~(\ref{eq:intmat})), while for VE/VdE we got a simpler Franck-Condon overlap according to Eq.~(\ref{eq:Vwv}) between two bound vibrational levels of the target.

This can be once again seen in Figs.~\ref{fig:8} and~\ref{fig:9}, where the DR, DE and the $\Delta v^+=\pm 1$ excitation (VE and VdE) rate coefficients of BeH$^{+}$ for $v_{i}=0,1,2$ are plotted. By increasing the initial vibrational level of the target,  we get larger rate coefficients for both dissociative processes. On the contrary, the VE and VdE rate coefficients are not showing this tendency since the cross section depend more strongly on the initial vibrational level of the target.  

\section{Conclusions}{\label{sec:conclusions}}

The present paper extends our previous studies~\cite{Niyonzima2013, Niyonzima, bed, Pop2021} of the BeH$^{+}$ and its isotopologues on the reactive collisions with electrons by considering higher collision energies facilitating the study of the dissociative excitation, occurring above the dissociation threshold of the molecular ion target. 
Starting from the original molecular data sets~\cite{Niyonzima2013}, and using simple scaling laws we have enlarged our set of molecular states participating to the fragmentation dynamics, resulting in a more realistic description of the dissociation and excitation of the molecular target. 
We produced cross sections and rate coefficients for DR, DE, VE, and VdE processes up to 12 eV collision energy, useful for the detailed kinetics modeling of the cold plasma close to the wall of the fusion devices.

The data underlying this article will be shared on reasonable request to the corresponding author.

\section*{Acknowledgements}
The authors acknowledge support from F\'ed\'eration de
Recherche Fusion par Confinement Magn\'etique (CNRS and
CEA), La R\'egion Normandie, FEDER, and LabEx EMC3
via the projects PTOLEMEE, Bioengine COMUE Normandie Universit\'e, the Institute for Energy, Propulsion and Environment (FR-IEPE), from the l’Agence Universitaire de la Francophonie en Europe Centrale et Orientale (AUF ECO) via the project CE/MB/045/2021 "CiCaM – ITER” and the European Union via COST(European Cooperation in Science and Technology) actions: TUMIEE (CA17126), MW-Gaia (CA18104),  MD-GAS(CA18212), PROBONO (CA21128), PhoBioS(CA21159), COSY(CA21101) and DYNALIFE(CA21169). The authors are indebted to Agence Nationale de la Recherche (ANR) via the project MONA. This work was supported by the
Programme National “Physique et Chimie du Milieu Interstellaire” (PCMI) of CNRS/INSU with INC/INP co-funded by CEA and CNES. J.Zs.M. thanks the financial support of the National Research, Development and Innovation Fund of Hungary, under the FK 19 funding scheme with project no. FK 132989. KC thanks SERB, India, for financial support under grant no. CRG/2021/000357.

\end{document}